\documentclass[preprint,aps,showpacs]{revtex4}
\usepackage{epsfig}
\usepackage{float}
\RequirePackage{amssymb,amsmath}
\usepackage[latin1]{inputenc}
\usepackage{graphicx}
\usepackage{indentfirst}
\usepackage{color}
\begin{document}

\title{Photo-fission of  $^{232}$Th and $^{238}$U  at intermediate energies}
\author{A. Deppman$^a$, E. Andrade-II$^a$, V. Guimar\~aes$^a$, G. S. 
Karapetyan$^a$ and  N. A. Demekhina$^b$.}

\affiliation{a) Instituto de Fisica, Universidade de Sao Paulo \\ 
Travessa R da Rua do Matao, 187, 05508-900 S~ao Paulo, SP, Brazil \\ 
b) Yerevan Physics Institute, Alikhanyan Brothers 2, Yerevan 0036, Armenia\\
Joint Institute for Nuclear Research (JINR), Flerov Laboratory of Nuclear 
Reactions (LNR), Joliot-Curie 6, Dubna 141980, Moscow region Russia}

\begin{abstract} 
In this work we present an analysis of the yields of fission fragments 
induced by bremsstrahlung photons with endpoint energies of 50 and 3500 MeV 
on $^{232}$Th and $^{238}$U targets using the simulation code CRISP. 
A multimodal fission option was added to this code and an extension 
of the calculation to the properties of the fission products is presented. 
By dividing the fissioning nuclei according to their fissionability, 
an approach is introduced which accounts for the contribution of symmetric 
and asymmetric fission. By adopting this procedure, it was possible
to calculate the main parameters for the fission fragment charge distribution
such as the most probable charge for a given fission product mass chain and its 
corresponding width parameter. Also, it was possible to reproduce features 
of fragment mass distribution and evaluate the fissility of fissioning 
nuclei for photon-induced fission of  $^{232}$Th and $^{238}$U.  
\end{abstract}
\pacs{25.85.Jg}

\maketitle

\section{Introduction}
\indent 
Despite almost seventy years of investigations on nuclear fission, this 
process still continues to be of great interest. 
The disintegration of the nucleus into 
two fragments of similar masses is accompanied by a complete rearrangement 
of the nuclear structure, and the dynamical process leading to fission 
determines the characteristics of the fragments in the final states.
Investigation of the photo-fission process in heavy nuclei is very interesting 
not only for the study of the fission mechanism itself, but also to obtain 
information about total photo-absorption \cite{bia93,sana00,ceti00}. 
Photons provide a convenient tool to study nuclear properties of a fissile 
system and to explore nuclear transformations 
at large deformations. In the case of photo-nuclear reactions, volume 
absorption dominates and, as a result, the photon effectively  ``heats'' the 
nucleus. For this reason, the excitation of the nucleus in different energy 
ranges reflects the nature of the interaction with incident photons. 
In the analysis of experiments with bremsstrahlung photons, the calculations 
of the total photo-fission yield takes into account the contributions of the 
different interaction modes by summing over the entire spectrum of photons. 

The calculation of fission cross section within different models and their 
comparison with data provides an opportunity to estimate the validity of the 
various photo-absorption mechanisms as well as to investigate 
characteristics of the processes taking place in reactions induced by 
different probes. Among the properties that can be used to compare 
calculations with data are charge and  mass distributions of the fragments, 
energy dependence of the fission cross section, and the ratio of 
symmetric and asymmetric components of fission products. 
In the present work, calculations for fragments produced in photo-fission of 
heavy nuclei, $^{232}$Th and $^{238}$U, induced by bremsstrahlung photons
with endpoint energies of 50 and 3500 MeV are presented. The most important 
quantities calculated are mass- and charge-distributions for each of the 
target nuclei at the two energies, 50 and 3500 MeV. 
Comparison with data is performed and the analysis allows the 
extraction of information about the fission process following the absorption of 
the photon. The data considered here were obtained from experiments described 
in detail elsewhere \cite{Nina1,Nina2} and the calculations were performed with 
the CRISP code \cite{Deppman2004}.

\section{The CRISP code}
\indent
CRISP is a Monte Carlo code which simulates, in  a two step process,
nuclear reactions induced by photons or protons \cite{Deppman2004}. 
 First, an intranuclear cascade is simulated following a time-ordered 
sequence of collisions in a many-body system \cite{Kodama,Goncalves1997}, 
and when the intranuclear cascade finishes, the evaporation of nucleons and 
alpha-particles begins, in competition with the fission process 
\cite{Deppman2001}. In the simulation, reactions can be initiated by 
intermediate and high energy protons \cite{Goncalves1997} or 
photons \cite{Pina,Deppman2002a,Deppman2002b}. The CRISP code has been shown 
to be reliable in reproducing photon induced reactions and gave good results 
for total photo-nuclear absorption cross sections for energies from 
approximately 50 MeV, where the quasi-deuteron absorption mechanism is 
dominant, up to 3.5 GeV, where the so-called photon-hadronization mechanism  
leads to a shadowing effect in the cross section \cite{Deppman2006}. 

One important feature of the code, in simulating the intranuclear cascade,
 is the Pauli blocking mechanism, which avoids violation of the Pauli 
principle. In the CRISP code a strict verification of this principle is 
performed at each step of the cascade, resulting in a more realistic 
simulation of the process. The advantages of such an approach have been 
discussed elsewhere (see for instance ref. \cite{Deppman2004} and references 
therein). In the evaporation/fission competition that follows the intranuclear 
cascade, Weisskopf's model is adopted to calculate the branching ratios of the 
evaporating channels, which includes evaporation of neutrons, protons and 
alpha-particles \cite{Deppman2001,Deppman2002a,Deppman2002b} and 
the Bohr-Wheeler-model is adopted for fission.  When one particle is emitted 
during the evaporation process, the excitation energy of the final state 
nucleus is calculated by $E_x^{(f)}=E_x^{(i)}-(B+V+\varepsilon)$, where 
$E_x^{(f)}$ and $E_x^{(i)}$ are the excitation energy of the final and 
initial nucleus, respectively, $B$ is the evaporated particle separation 
energy, $V$ is its Coulomb potential, and $\varepsilon$ is the mean kinetic 
energy of the emitted particle, which is fixed at 2 MeV.
The code has provided good agreement of photo-fission cross section data 
\cite{Deppman2004}. The CRISP code has also already been used to evaluate 
mass distributions of fragments for fission induced by photons at 
intermediate energies \cite{AndradeII2009}, and to calculate spallation 
yields and neutron multiplicities for reactions induced by high energy protons 
\cite{Anefalos2008}, giving results in good agreement with data. Moreover, the 
code has already been used in studies of the ADS (Accelerator Driving System) 
nuclear-reactors \cite{Anefalos2008, Anefalos2005a, Anefalos2005b, Mongelli}.

\section{Description of the Model and the Calculations}

The fission process has been successfully described by the Multimodal - 
Random Neck Rupture Model (MM-RNRM) \cite{Brosa1990}, which takes into 
account the collective effects of nuclear deformation during fission 
through a liquid-drop model, and includes single-particle effects through 
microscopic shell-model corrections. The microscopic corrections create 
valleys in the space of elongation and mass number, each valley corresponding 
to a different fission mode \cite{Brosa1990}. 
In the MM-RNRM, the yield of fragments is described for each mode by a Gaussian 
distribution which is characterized by the fragment's mass and atomic number, 
$A$ and $Z$, respectively. According to this model, the 
mass-yield curve can be decomposed into three distinct fission components: 
one symmetric Superlong and two asymmetric, Standard I and Standard II. 
Superlong mode fragments are strongly elongated with masses around $A_{f}$/2. 
The Standard I mode is characterized by the influence of the spherical neutron 
shell $N_{H}\sim$82 and proton shell $Z_{H}\sim$50 in the heavy fragments 
with masses $M_{H}\sim$132-134. The Standard II mode is characterized by 
the influence of the deformed neutron shell closure $N$=86-88 and proton shell 
$Z_{H}\sim$52 in the heavy fragments with masses $M_{H}\sim$138-140.
There is a new investigation of the influence of shell closures by 
Schmidt and Jurado \cite{schmidt12} indicating that the fission of a large 
number of isotopes is governed not by the neutron closed or deformed shells 
at N=82 or N=88 but by the proton number $Z=54$. 
Here we consider that fission can take place through one of these three 
different modes, and that the total mass-yield is obtained by the sum of the 
three Gaussian functions \cite{younes01}:

\begin{align}
 \begin{split}
  Y_A=&
\frac{1}{\sqrt{2\pi}}\bigg[\frac{K_{1AS}}{\sigma_{1AS}}
\exp\left(-\frac{(A-A_S-D_{1AS})^2}{2\sigma^2_{1AS}}\right)+
\frac{K'_{1AS}}{\sigma'_{1AS}}\exp\left(-\frac{(A-A_S+D_{1AS})^{2}}
{2\sigma'^2_{1AS}}\right)+\\
&\frac{K_{2AS}}{\sigma_{2AS}}\exp\left({-\frac{(A-A_S-D_{2AS})^2}
{2\sigma^2_{2AS}}}\right)+
\frac{K'_{2AS}}{\sigma'_{2AS}}\exp\left({-\frac{(A-A_S+D_{2AS})^2}
{2\sigma'^2_{2AS}}}\right)+\\
	&\frac{K_S}{\sigma_S}\exp\left({-\frac{(A-A_S)^2}
{2\sigma^2_S}}\right)
\bigg],
 \end{split}
\end{align}
where $A$ is the fragment mass number; $A_S$ is the mean mass number 
which determines the center of the Gaussian functions; and $K_i$, $\sigma_i$, 
and D$_i$ are the contribution, dispersion and position parameters of 
the $i^{th}$ Gaussian functions. The indexes $AS$ and $S$ designate the 
asymmetric and symmetric components.
Note that, when using the CRISP code, it is possible to work on 
event-by-event basis, and therefore the parameter $A_S$ in Eq. (1) is 
completely determined by the mass of the fissioning nucleus, 
that is, $A_S=A_{ff}/2$.
During the calculation, we have defined two types 
of fissioning nuclei, before ($A_f$) and after ($A_{ff}$) post-fission 
neutron evaporation, where A$_H$+A$_L$=$A_{ff}$, and the indexes
$H$ and $L$ stand for heavy and light mass fragments.
The quantities  $A_S$ + D$_{iAS}$ = A$_H$ and 
$A_S$ - D$_{iAS}$ = A$_L$ determine the positions of the heavy and 
light peaks of the asymmetric components on the mass scale. 
The values of A$_H$ + A$_L$ and 2$A_S$ were treated as the masses of 
nuclei that undergo fission in the respective channel. 
Typical values for these parameters obtained in a fitting process can be 
found in refs.  \cite{Nina1} and \cite{Nina2}.

The fragment charge distribution can be estimated by considering the 
Gaussian functions in the form \cite{kudo,duijvestijn}: 

\begin{eqnarray}
Y_{A,Z}=\frac{Y_A}{\Gamma_z\pi^{1/2}}\exp\left({-\frac{(Z-Z_p)^2}{\Gamma_z^2}}\right),
\label{frag_yield}
\end{eqnarray}

\noindent 
where $Y_{A,Z}$ is the independent yield of the nuclide ($Z,A$), $Y_A$ is 
the total yield for a given mass number $A$, $Z_p$ is the most probable charge 
for isobars with mass number $A$ and $\Gamma_z$ is the width parameter. 
The parameters $Z_p$ and $\Gamma_z$ can be represented as 
slowly varying linear functions of the mass numbers of fission fragments:

\begin{eqnarray}
Z_p=\mu_1+\mu_2A\,,
\label{zp}
\end{eqnarray}
and
\begin{eqnarray}
\Gamma_{z}=\gamma_1+\gamma_2A\,,
\label{gp}
\end{eqnarray}

 The multimodal model  has been used previously to 
describe spontaneous fission \cite{Wagemans}, low-energy induced 
fission \cite{Bockstiegel2008}, fission induced by thermal-neutrons 
\cite{Hambsch2002, Hambsch2003, Hambsch1989} and 12 MeV protons 
\cite{ohtsuki89},
and even fission induced by intermediate energy probes such as 190 MeV 
protons \cite{duijvestijn}, neutrons at energies up to 200 MeV 
\cite{maslov03}, and also by heavy-ions \cite{Itkis,Pokrovsky}. 

The CRISP code was then adapted to consider the multimodal model by the 
use of Eqs. (1) and (2).  To determine the fission fragment masses by 
the CRISP code it is  necessary to  attribute values for the parameters used 
in the multimode approach,  which is not a trivial problem.
Since the CRISP code simulates the entire process up to the point of fission, 
the fissioning nucleus of all events is known, which leads to the fact 
that $A_S$ cannot be taken as a free parameter but as a 
distribution instead. Therefore, at every point of decision on the fission 
channel, the appropriate value for $A_S$ is used considering 
the nucleus which is undergoing fission.  
Whenever the fission channel is chosen, the masses and atomic numbers of the 
heavy fragments produced, $A_H$ and $Z_H$, respectively, are sorted 
according to Eq. (1). The light fragments are obtained according to 
$A_L=A_f-A_H$ and $Z_L=Z_f-Z_H$, where $A_f$ and $Z_f$ are the 
mass and atomic number of the fissioning nucleus, respectively.
As mentioned, $A_S$  is the average mass number of the fragments 
for the symmetric component, which determines the center of 
the Gaussian function. It is related to the mass of fissioning nucleus by
2$A_S$=$A_f$, where $A_f$=$A_H$+$A_L$.

As a final step, all fragments, obtained in the above fashion, evaporate 
according to the model of evaporation/fission competition already mentioned. 
The energy of each fragment is determined using:

\begin{align}
 E_i = \dfrac{A_i}{A_f} E_{frag},
\end{align}
where $E_{i}$ and $A_{i}$ are the excitation energy and the mass number of the 
fragment $i$. $E_{frag}$ is the total excitation energy of the fragments which 
is assumed, as an approximation, to be equal to the excitation energy of the 
fissioning system.

\section{Results and Discussion}

The CRISP code is used here to analyse data on fission fragment distributions
produced by photo-fission of $^{232}$Th and $^{238}$U. The data were obtained 
from radiation of  bremsstrahlung photons on  $^{232}$Th and $^{238}$U targets.
The bremsstrahlung photons with endpoint energies of 3500 and 50 MeV
were obtained with the Yerevan electron synchrotron and a linear accelerator 
of injector type, respectively. In these measurements, the electrons were 
converted into a bremsstrahlung photon beam by means of a tungsten converter 
of about 300$\mu$m. The photon-beam intensity was determined with a Wilson 
type quantameter, and the obtained average values were $10^{11}$ and $10^9$ 
equivalent photons per second at the maximum energy of 3500 MeV and 50 MeV, 
respectively. The $^{232}$Th and 
$^{238}$U targets had 20$\mu$m and 70$\mu$m of thickness, respectively. 
The yields of radioactive fission fragments were measured in off-line mode 
with a HpGe semiconductor detector. The energies of the gamma 
transitions and the relations between their intensities, as well as 
half-lives, were used to select reaction products and to determine their 
yields. Measurements of the gamma spectra started about 120 minutes after 
the irradiation was finished and lasted for a year. More details on 
these experiments can be found in refs. \cite{Nina1} and \cite{Nina2}.

Usually, in such off-line analysis, the detected fragments  represent the 
final products of the fission process after neutrons and gamma-rays are
emitted from both the fissioning nucleus and the primary fragments. 
Thus, comparison between calculation and experiment allows the unfolding 
of the contributions of pre- and post-scission neutron emissions. 
Here we are considering the charge and mass characteristics of the 
fission products to obtain information about the hot nuclear system 
and its decay channels.

\subsection{Charge Distribution}

As discussed above, according to Eq.~(\ref{frag_yield}), the charge 
distribution for an isobar chain with mass number $A$, from a fissioning heavy 
nuclei, is characterized by a Gaussian shape with parameters,  $Z_p$ and 
$\Gamma_z$, where  $Z_p$ and $\Gamma_z$ are the most-probable charge and 
the corresponding width of the distribution. The best representation 
regarding the most probable charge $Z_p$ is a  linear function of the mass 
of the fission fragment, as given by the empirical Eq. (3) \cite{bra77}.
Although these parameters have a linear dependence on A, through 
Eqs.~(\ref{zp}) and~(\ref{gp}), respectively, experimentally, the $\Gamma_z$ 
are practically independent of A \cite{kudo,morrissey80,bra77}.

In Table I, we summarized the experimental and calculated
relevant parameters, $\mu_1$,  $\mu_2$, $\gamma_1$ and  $\gamma_1$ used
in the present work to determine  $Z_p$ and $\Gamma_z$.
As one can see, the calculated parameter $\gamma_2$ is very small for all
four cases, indicating that $\Gamma_z$ is almost constant. 
Experimentally, instead of considering the  $\Gamma_z$, we averaged 
the $FWHM$ (Full Width at Half Maximum) of all charge distributions. 
The small deviation around the average values is another indication that 
this parameter is practically independent of A. 
Although it has been experimentally observed that
 the width of charge distribution $\Gamma_z$ is independent of A, 
there is a small mass dependence in the calculated values 
given by the parameter $\gamma_2$ in Eq. (3) and shown in Table 1. 
From the value of FWHM and $\gamma_2$ in Table 1, it is clear, that as 
the excitation energy increases, the charge distribution becomes slightly 
wider. In general, a similar dependence was observed in fission induced by 
particles of different types at excitation energies up to 
100 MeV \cite{chung82}.

In Fig. 1 we plot the difference between the calculated and experimental values 
for  $\Gamma_{z}$ of the charge distributions for the $^{238}$U and $^{232}$Th 
targets at the two endpoint energies. It is possible to observe that the 
overall difference is around $\pm$0.3 units, corresponding to approximately 
25\% of the total width $\Gamma_z$, and that the difference is in the same 
range for both endpoint energies. The deviation between calculation and 
experiment presents an approximately linear increase with mass number $A$ 
for all four cases studied, going from  -0.3 for $A \approx$ 85 
to  +0.3 for $A\approx$ 150. The fluctuation observed in this deviation 
seems to be a manifestation of fragment shell effect ($Z_p\sim$50), 
which might not have been completely taken into account by the linear 
parametrization of Eqs.~(\ref{zp}) and~(\ref{gp}). This effect 
can be  clearly seen in the direct comparison of calculated and 
experimental most-probable charge, $Z_p$ shown in Figs. 2 and 3 for the 
$^{238}$U and $^{232}$Th targets, respectively. Figs. 2 and 3 show both the 
experimental values of the most probable charge $Z_p$ and the values 
calculated by the CRISP model for $^{238}$U  and $^{232}$Th, respectively, 
at the two incident energies. In general, the experimental 
values of $Z_{p}$ cogently  fit to those obtained by model calculations. 
To be more quantitative, the reduced $\chi^{2}$ was used to evaluate 
the comparison  between data and CRISP calculations. The results obtained are, 
for $^{238}$U,  $\chi^{2}=0.80$ at $E_{max}=50$ MeV and $\chi^{2}=0.54$ 
for $E_{max}=3500$ MeV; and for $^{232}$Th: $\chi^{2}=0.66$ for $E_{max}=50$ MeV; 
$\chi^{2}=0.38$ at $E_{max}=3500$ MeV). 
It can be noticed that both calculated values and data present an overall 
linear 
trend with the mass number $A$, as given by the parameters $\mu_1$ and $\mu_2$.

As one can see in Table 1, the value of the  calculated parameters $\mu_2$ 
for the uranium isotope, at both energies, are slightly higher than the 
experimental ones. As a result, the values obtained for the most 
probable charge $Z_p$, for a given mass number, are shifted to a larger value 
(more neutron deficient). In Figs. 2 and 3 one can see that the calculated 
values for $Z_p$ are mostly on the left side, proton-rich side, of the 
experimental values, especially for low energy fission for both isotopes. 
This is a reflection of the dependence of the parameter $\mu_2$ on charge 
and mass of the fissioning nucleus, and thus, on the average number of emitted 
neutrons from the fission fragments \cite{kudo}. The 
calculated values for the parameter $\mu_1$ are about one unit below the 
experimental values for the uranium target at both energies and one unit 
above for the thorium target. The higher measured values for $\mu_1$, 
in comparison with the calculated ones, indicate that the calculation favors 
a most probable charge for the fission products of uranium target closer to 
the stable nucleus. With the increase of excitation energy and the atomic 
number of the fissioning nucleus, one would expect that $Z_p$ of an isobaric 
chain would become closer to the Z values of a more stable nucleus. 
Therefore, the model is considering greater evaporation of neutrons from the 
fission fragments. In the case of thorium the situation is the opposite, 
but the agreement with calculated values is better. We can assert that 
the model considers more correctly the evaporation of neutrons for the Th 
target.
Also, as one can see in both Figs. 2 and 3, for $A$ around 125, the data show  
a slight deviation between calculated and experimental 
$Z_p$ values in the region of $A>$130.  This deviation is more prominent in 
the case of the $^{238}$U target. These results suggest that the CRISP 
calculation
can be improved by including shell effects in the parametrization of $Z_p$.  
However, aside from the slight deviation discussed above, the calculation 
seems to present a general small shift to neutron deficient fragments, 
which also can be corrected in the next upgraded version of the code.

This analysis of parameter values shows that, in general, the CRISP model 
describes fairly well the most probable charge of the charge distribution for 
the fission fragments for $^{232}$Th and for $^{238}$U, with 
some fluctuations in $\Gamma_z$. It can, however, be 
assumed that the model takes into account the main properties of fissile 
systems and provide a way to predict some characteristics of the fragment 
charge distribution.

As can be observed in the data, as we increase the endpoint energy the 
composite system has a higher excitation energy producing, on average,
fissioning nuclei with less neutrons due to the higher number of evaporated 
neutrons (proton evaporation is suppressed due to the Coulomb barrier). 
As a consequence, the following produced heavier fragments will also evaporate 
more neutrons and as a result, the fragments with same A will have larger 
$Z$ shifting, on average, the $Z_p$ parameter to higher values. 
 Similar behavior is observed for fission induced by different probes such 
protons and neutrons on $^{238}$U and $^{232}$Th 
\cite{kudo,chung81,nethway72,mchugh68}, indicating that the charge 
distributions are determined more by the excitation energy and nuclear 
properties of the reaction products than by the choice of projectile.

\subsection{ Mass Distribution}

The origin of asymmetric fission is associated with the shell structure 
of the fissioning nucleus and nuclear fragments \cite{Strutinsky}, whereas 
symmetric fission is consistent with a classical liquid-drop model of the 
fissioning nucleus \cite{Myers}, and it is the most relevant mechanism for 
fission of highly excited ($>$ 50 MeV) nuclei. Therefore, mass distributions 
of fragments depend on the mass of the fissioning nuclei and on its 
excitation energy.
 
The calculations with the CRISP code consider the three-mode hypothesis 
discussed previouly, corresponding to one symmetric (Superlong) and two 
asymmetric dynamics (Standard I and Standard II). The results obtained allow 
a  complete analysis of the fragment-mass distributions for $^{238}$U and 
$^{232}$Th. The total fission mass-yield distribution as a function of the 
product mass number ($A$) for $^{238}$U and  $^{232}$Th targets  at 
$E_{max}=50$, 3500 MeV are shown in Fig. 4 (a,b) and  Fig. 5 (a,b), 
respectively. 
Qualitatively, the agreement between the calculation and the experimental 
data is fairly good at both endpoint energies,  particularly for low energies,
 where the calculations reproduce the experimental data quite well. 
This fact shows that the model correctly takes into account the influence of 
shell effects for low energy fission, associated with asymmetric fission modes: 
the strong spherical neutron shell at $N=82$ and the deformed neutron shell 
at $N=86-88$ become dominant and lead to asymmetric fission. 
It is also possible to observe in Figs. 4 and 5 that a better agreement 
between calculation and data is achieved for the lower endpoint energy. 
This can be due to the fact that the parameters used in Eq. (1) are based 
on a systematic analysis for low energy fission, which may not be as good for 
higher energies. To a lesser extent, by comparing the results for $^{232}$Th 
with those for $^{238}$U, one can observe that for  uranium the agreement 
with data is better for both endpoint energies. This also can be due to the 
fact that the systematics used to set the parameters was done for $A>220$, 
and the results extrapolated to lower mass. It is possible to conclude that 
the values for $D$ and $\Gamma$ for $A<220$ are somewhat overestimated.

Another parameter that can be used to compare data and calculations is 
the integrated total fission yield $Y_F$, given by:

\begin{equation}
 Y_F=\frac{1}{2}\int Y_A(A) dA\,.
\end{equation}

The result of this calculation together with data on total fission
yields in units of mb per equivalent quanta (mb/eq.q) as well as 
some other calculated quantities and their corresponding experimental
values are listed  in Table II for comparison. 

At intermediate energies the discrepancy between experiment and calculation 
becomes larger for all observed parameters, except for total fission yield 
values. The reason for this can be due to the character of bremsstrahlung. 
As the later has a continuous spectrum, the measurements at high energy 
include also low-energy photons. From the calculated mass distribution it
 can be concluded that the model underestimates the low-energy part of the
bremsstrahlung spectrum. The calculated mean mass number of distributions after 
evaporation of post-scission neutrons ($A_{ff}$)$_{cal}$ is shifted to lower 
masses in comparison to the experimental ones with increasing energy. It 
means that the yields of symmetric fission fragments grow faster, because of 
the considerable amount of high-energy photons.

Moreover, we also calculated the mass-yield distribution of the fissioning 
nuclei, before they undergo fission, which are shown in Fig. 6.
This figure allows the comparison of the fissioning system 
distributions for thorium and uranium targets at both endpoint energies. 
These distributions result from a complex balance between evaporation and 
fission processes. Apart from the obvious fact that  there are no 
fissioning nucleus with mass $A>232$ for the case of  $^{232}$Th, the 
fissioning nucleus distribution extends much below the range of those systems 
in the case of $^{238}$U. It is clear that low mass nucleus fragments are more 
abundant in the case of thorium fission distribution than in the case of 
uranium. One of the consequences of the predominance of low mass nuclei in 
the evaporation chain for $^{232}$Th is a relatively lower peak to valley
ratio, $P/V$, with respect to that of uranium. Another important concern is 
that the uncertainties due to the extrapolation of the values for the 
parameters in Eq. (1), which is based on the systematic analysis for $A>220$, 
are larger in the case of thorium than in the case of uranium. These facts 
together can explain why the results presented here for $^{238}$U are in better 
agreement with data than for $^{232}$Th.

The nature of the interaction of photons and hadrons with nuclei is very 
different but, the fission process, as mentioned, depends mostly on the 
excitation of the fissioning nucleus. Thus, to compare data of fission 
induced by different particles it is useful to use systems with equivalent 
excitation energy.  Our result of fission induced by photons with endpoint 
energy of 50 MeV on $^{238}$U and $^{232}$Th are, thus,  compared with fission 
induced by low energy protons, neutrons and alpha particles on the same 
targets \cite{ohtsuki89,younes01,mathies90}. The main characteristics of 
mass distributions such as widths of different fission modes, position 
parameter $D$ of asymmetric modes and the average mass number of fragments 
for the symmetric component, $A_S$, are in good agreement among these data, 
giving a clear indication that it is not the incident particle  but the 
structure of the fissioning nucleus and the characteristics of the 
fragments which determine the mass distribution. A detailed calculation 
with a statistical model of multi-modal fission was 
performed for intermediate energy neutrons (200 MeV) by Maslov \cite{maslov03}. 
It was observed that the largest contribution was obtained 
for the total fission cross section events with the emission of 5-7 neutrons 
on average. This result is in good agreement with our photo-fission data 
for 3500 MeV photons (excitation energy around 120 MeV) on both  
$^{238}$U and $^{232}$Th targets, where the average number of prefission 
neutrons is 5. 
Another comparison that can be made with reactions induced by
different probes on $^{238}$U and $^{232}$Th is the peak to valley ratio (P/V) 
of the mass distribution, which is an indirect way to interpret symmetric and
 asymmetric mode contributions  for fission. The experimental values 
obtained in the present work together with data obtained for fission induced by 
photons \cite{schroder70,karamian00}, protons \cite{chung81,kudo82,chung82} and 
neutrons \cite{younes01,maslov03,Gobachev76} on the same targets are presented 
in Fig. 7.  As can be observed in the figure, there is a general trend 
for all data of different probes indicating again that it is the excitation 
energy which is more relevant for the fission process than the type of 
incident particles.

A systematization of cross sections for symmetric and asymmetric fission in 
a wide range of nuclei, carried out by Chung {\it et al.}  \cite{chung82} 
showed that it is possible to use an empirical expression to estimate the 
probability of the different 
fission modes. In order to characterize this factor quantitatively, 
Chung {\it et al.}  have introduced a critical value of the fissility 
parameter, in the form:
\begin{eqnarray}
(Z^{2}/A)_{cr.}=35.5+0.4(Z-90),
\label{critical_fission_parameter}
\end{eqnarray}

\noindent 
where Z, and A are the atomic number and mass of the fissioning nucleus.

Thus, symmetry of the fission fragment mass distribution is given, 
by the ratio $Z^{2}/A$.  For nuclei with $Z^{2}/A$  greater than the 
critical value, given by Eq.~(\ref{critical_fission_parameter}), the 
symmetric fission mode is dominant, while for values below  the critical 
fissility parameter, the fission dynamics led predominantly to asymmetric
 fragment distributions. For  $^{238}$U, the parameter 
($Z^{2}/A$)$_{cr.}$ is 36.3, and for $^{232}$Th it is 35.5. Thus, at low 
energy (with an average of no more than three evaporated neutrons) for 
$^{238}$U it is natural to expect predominantly asymmetric fission. On the 
other hand, it is well-known that the symmetric component of the fission 
process increases as the excitation energy of the fissioning nucleus 
increases. This can be roughly understood as an effect of two factors. 
The first one is that, with increasing energy, shell effects become less 
pronounced and therefore, fission tends to be predominantly symmetric. The 
other factor is the length of the evaporation chain, which increases as the 
fissility decreases. Since the evaporation is dominated by neutron emission, 
longer evaporation chains lead to nuclei in the proton-rich side of the 
stability valley, and therefore the fissility parameter, $Z^2/A$, tends to be 
above the critical value, resulting in the predominance of symmetric fission.

The fissility of $^{232}$Th is lower than that for $^{238}$U, and then one can 
expect that the former nucleus presents a longer evaporation chain. In fact, 
this can be observed for instance, by the difference of the target nucleus 
$^{238}$U and the mean mass of the fissioning nucleus  after evaporation of 
post-scission neutrons,  $(A_{ff})_{cal}$, which for the endpoint photons of 
3500 MeV is ten mass units. The same difference is twelve mass  units in the 
case of thorium. This indicates that the evaporation chain is two steps 
longer in the case of thorium. Also, looking at the peak to valley ratio 
($P/V$), it is possible to observe that thorium presents a stronger 
contribution of symmetric fission when compared with that for uranium.

According to the well-known concept, the fissility is determined as the 
ratio of the fission yield and the yield of total photon absorption in a 
nucleus ($D$=$Y_{tot}$/$Y_{abs}$). In Fig. 8 we plot this fissility $D$
for  the $^{238}$U and $^{232}$Th targets, from the present work,
together with data for proton-induced fission of $^{241}$Am, $^{238}$U and 
$^{237}$Np nuclei \cite{Nina3,Nina4}, as a function of the fissility
parameter $Z^{2}/A$ of the fissioning nucleus. We also plot 
the estimated fissility by 
the CRISP code for photo-fission of $^{238}$U and $^{232}$Th. For the case of 
$^{238}$U  one can see that the fission probabilities are about the same, 
independently of the projectile used to excite the nuclear matter. 
The calculated fissilities for the thorium target are very close to the 
experimental values at both energies within the uncertainties. It should be 
mentioned that the $A$ used to determine the  $Z^{2}/A$ parameter is given by 
the $A_f$ listed in Table II, which are different for the calculated and 
experimental values. In the case of the uranium target the calculated 
values of the fissility at both energies overestimate the experimental data, 
especially  for the low energy  photons. This discrepancy may 
be due to a limitation of the model in taking into account 
all possible channels of decay of the excited nucleus being considered.

\section{Conclusion}

The present version of the CRISP code, which takes into account different 
channels of fission, was used to reproduce different aspects of photon-induced
 fission on actinides. Those calculations allow direct evaluation of the 
spectrum of fissioning nuclei. The comparison between calculated parameters
and data has shown that the calculations describe correctly the main 
characteristics of charge, such as the most probable charge for a given 
fission product mass chain and the width parameter for the photo-induced 
fission of $^{232}$Th and $^{238}$U targets at two very different energy regimes 
(bremsstrahlung photons with endpoint energies of 50 and 3500 MeV). The mass 
distribution of photo-fission fragments has been analyzed via the multimodal 
fission approach.  The results presented in this paper show fair agreement 
between calculation and experiment. The results of the calculations made it 
possible to determine the fissilities of the fissioning nuclei and compare 
them 
with those from other experiments. It was found that CRISP simulations better 
reproduce data for low-energy photon-induced fissioning systems.

\section*{Acknowledgment} G. Karapetyan is grateful to 
Funda\c c\~ao de Amparo \`a Pesquisa do Estado de S\~ao Paulo (FAPESP) 2011/00314-0, and to International Centre for Theoretical Physics (ICTP) under the 
Associate Grant Scheme. We thank prof. Wayne Seale for reviewing
the text.

\newpage

\begin{table}[!ht]\centering
\caption{Parameters used to determine experimental and
calculated charge distributions.}
\begin{tabular}{|c|c|c|c|c|}  \hline
Parameter&\multicolumn{2}{|c|}{$^{238}$U}&\multicolumn{2}{|c|}{$^{232}$Th}\\
\cline{2-5} &$E_{max}=50$ MeV&$E_{max}=3500$ MeV&$E_{max}=50$ MeV&$E_{max}=3500$ MeV\\
\hline ($\mu_1$)$_{exp}$ & 5.70$\pm$0.60 & 5.32$\pm$0.62 & 3.89$\pm$0.67 & 4.14$\pm$0.70\\
\hline ($\mu_2$)$_{exp}$& 0.356$\pm$0.005 & 0.362$\pm$0.005 & 0.371$\pm$0.005 & 0.356$\pm$0.005\\
\hline ($\mu_1$)$_{cal}$ & 4.10 & 4.10 & 5.00 & 5.00\\
\hline ($\mu_2$)$_{cal}$ & 0.380 & 0.380 & 0.370 & 0.370\\
\hline ($\gamma_1$)$_{cal}$&0.92&0.92&0.59&0.59\\
\hline ($\gamma_2$)$_{cal}$&0.003&0.003&0.005&0.005\\
\hline $FWHM$&1.03$\pm$0.12&1.09$\pm$0.13&1.13$\pm$0.14&1.14$\pm$0.15 \\ \hline
\end{tabular}
\end{table}

\newpage
\begin{table}[!ht]\centering
\caption{Calculated and experimental parameters obtained for the
mass distribution. The total fission yield $Y_F$; the position of the two 
peaks of asymmetric fission, $A_L$ and $A_H$; the mean 
mass of the mass distribution ($A_S$); mean mass of the 
fissioning nucleus (($A_f)_{cal}$) after evaporation of pre-scission neutrons 
from the compound nucleus; mean mass of the fissioning nucleus (($A_{ff})_{cal}$) 
after evaporation of post-scission neutrons from fragments; experimental mean
 mass of the fissioning nucleus (($A_f)_{exp}$), which includes both type of 
evaporated neutrons; the values of the peak-to-valley ratios ($P/V$). }
\begin{tabular}{|c|c|c|c|c|}  \hline
Parameter&\multicolumn{2}{|c|}{$^{238}$U}&\multicolumn{2}{|c|}{$^{232}$Th}\\
\cline{2-5} &$E_{max}=50$ MeV&$E_{max}=3500$ MeV&$E_{max}=50$ MeV&$E_{max}=3500$ 
MeV\\
\hline ($Y_F$)$_{exp}$ $(mb/eq.q)$ & 131$\pm$20&250$\pm$38&40$\pm$6&138$\pm$21\\
\hline ($Y_F$)$_{cal}$ $(mb/eq.q)$ & 149&268.8&39.6&129.6\\
\hline ($P/V$)$_{exp}$ & 11.4$\pm$1.7 & 2.16$\pm$0.40 & 7.9$\pm$1.6 & 0.84$\pm$0.17\\
\hline ($P/V$)$_{cal}$ &10.63 & 6.45 &4.80 & 1.45\\
\hline ($A_L$)$_{exp}$ & 98.0$\pm$1.9 & 97.0$\pm$1.9 & 91.5$\pm$1.8 & 94.0$\pm$1.8\\
\hline ($A_L$)$_{cal}$ & 98.0 & 95.0 & 94.0 &95.5\\
\hline ($A_H$)$_{exp}$ & 137.0$\pm$2.7 & 137.0$\pm$2.7 & 137.5$\pm$2.7 & 134.0$\pm$2.6\\
\hline ($A_H$)$_{cal}$ & 136.0 & 133.0 & 134.0 & 124.5\\
\hline ($A_S$)$_{exp}$ & 117.5$\pm$0.2 & 117.0$\pm$0.2 & 114.5$\pm$0.3 & 114.0$\pm$0.3\\
\hline ($A_S$)$_{cal}$ & 117.0 & 114.0 & 114.0 & 110.0\\
\hline ($A_f$)$_{exp}$ & 235.0 & 234.0 & 229.0 & 228.0\\
\hline ($A_f$)$_{cal}$ & 237.55 & 235.95 & 230.61 & 227.70\\
\hline ($A_{ff}$)$_{cal}$ & 234.0 & 228.0 & 228.0 & 220.0\\
\hline
\end{tabular}
\end{table}

\newpage
\begin{figure}
\epsfig{file=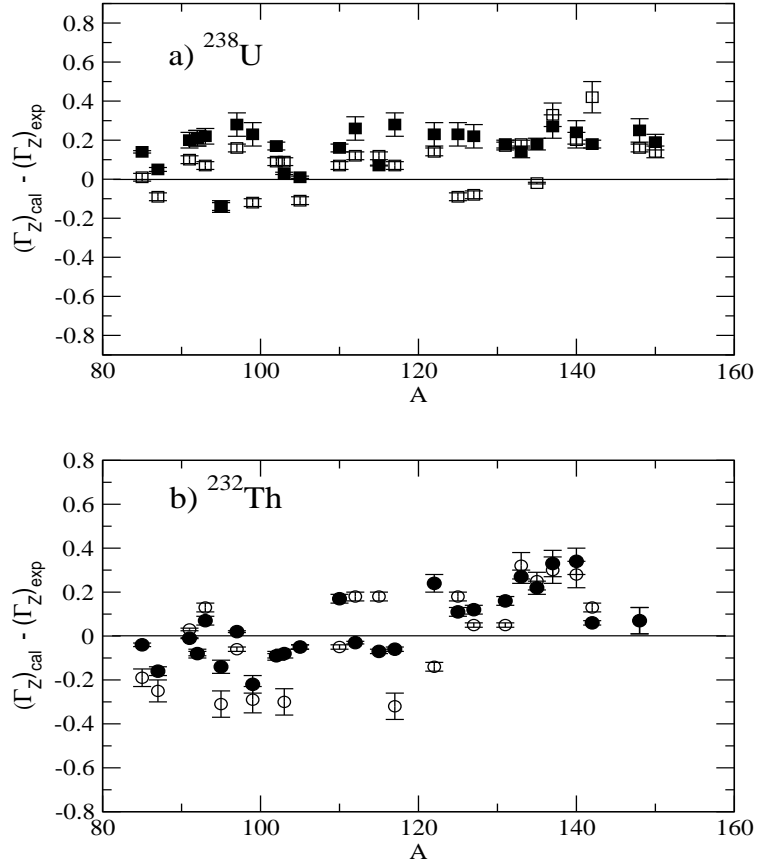,height=14cm,width=12cm,angle=0.}
 \caption{ The difference between the experimental and 
calculated values by the CRISP code  for the width of the charge distribution 
for a) $^{238}$U  and b) $^{232}$Th targets.
The open square and open circle symbols are for data taken at 
  $E_{max}=50$ MeV while the solid square and solid circle symbols were taken 
at  $E_{max}=3500$ MeV.}
\end{figure}

\begin{figure}
\epsfig{file=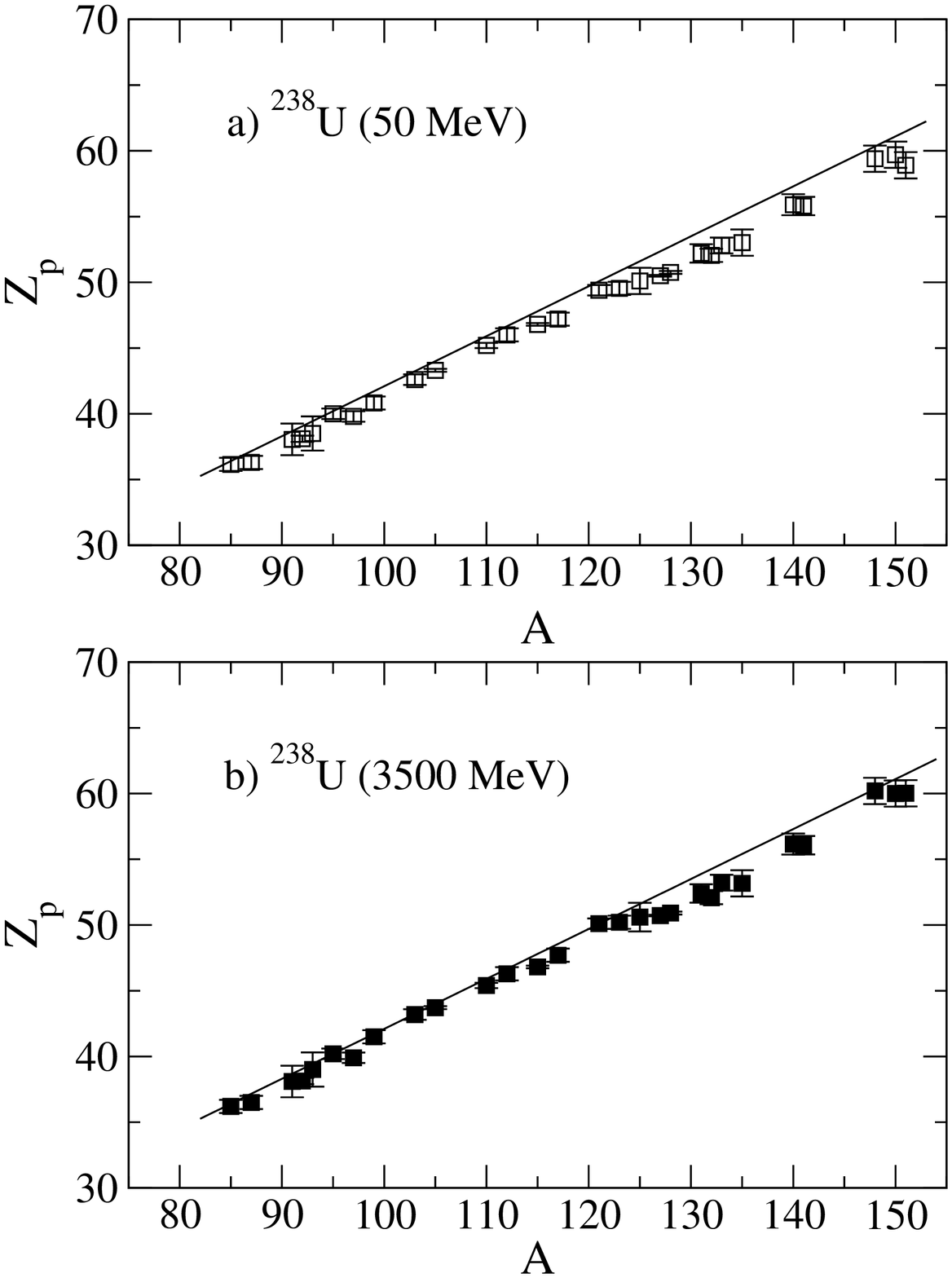,height=14cm,width=12cm,angle=0.}
\caption{The most probable charge $Z_{p}$ for induced photo-fission on 
a $^{238}$U target at bremsstrahlung photon endpoint energies of a) 
50 and b) 3500 MeV, respectively. The black solid lines are the results 
of calculations by CRISP.}
\end{figure}

\begin{figure}
\epsfig{file=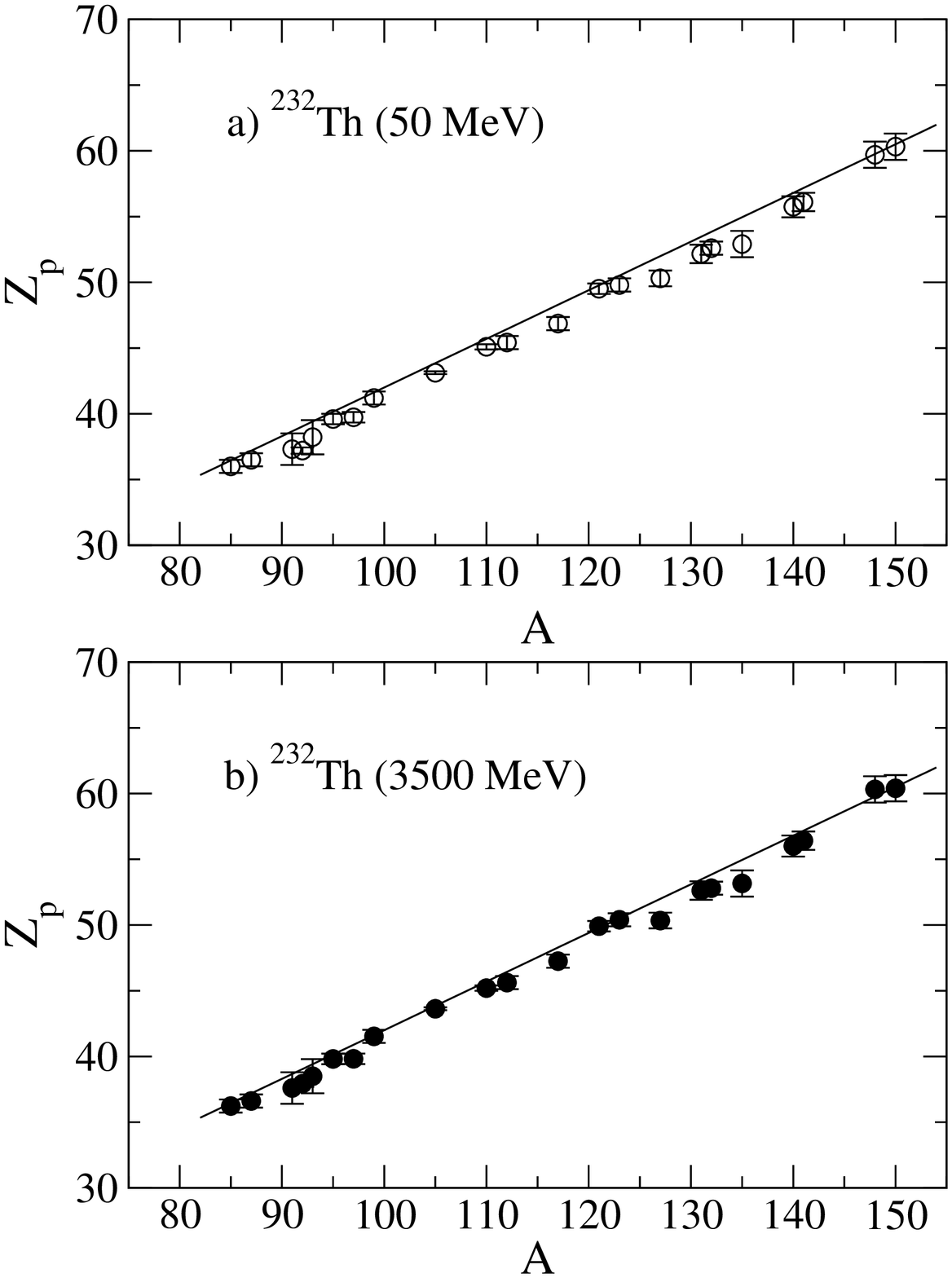,height=14cm,width=12cm,angle=0.}
\caption{The most probable charge $Z_{p}$ for induced photo-fission on
a $^{232}$Th target at bremsstrahlung photon endpoint energies of a) 
50 and b) 3500 MeV, respectively. The black solid lines are the results 
of calculations by CRISP.}
  \end{figure}

\begin{figure}
  \epsfig{file=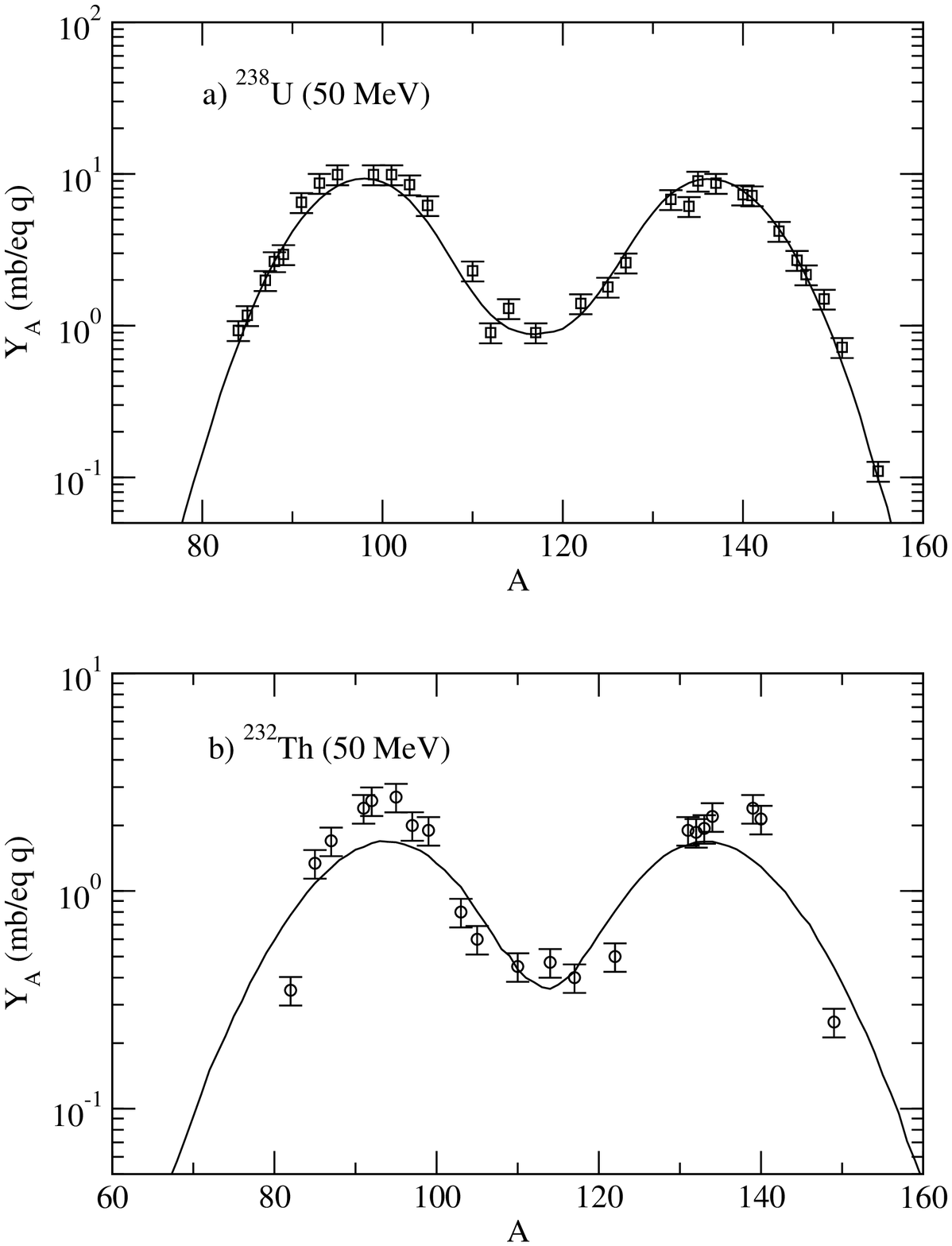,height=14cm,width=12cm,angle=0.}
  \caption{Mass distribution, mass-yield as a function of mass number
in units of mb per equivalent quanta (mb/eq q),
at $E_{max}=50$ MeV for a) $^{238}$U and b) $^{232}$Th targets. 
The solid line is the result of CRISP calculation.}
\end{figure}

\begin{figure}
  \epsfig{file=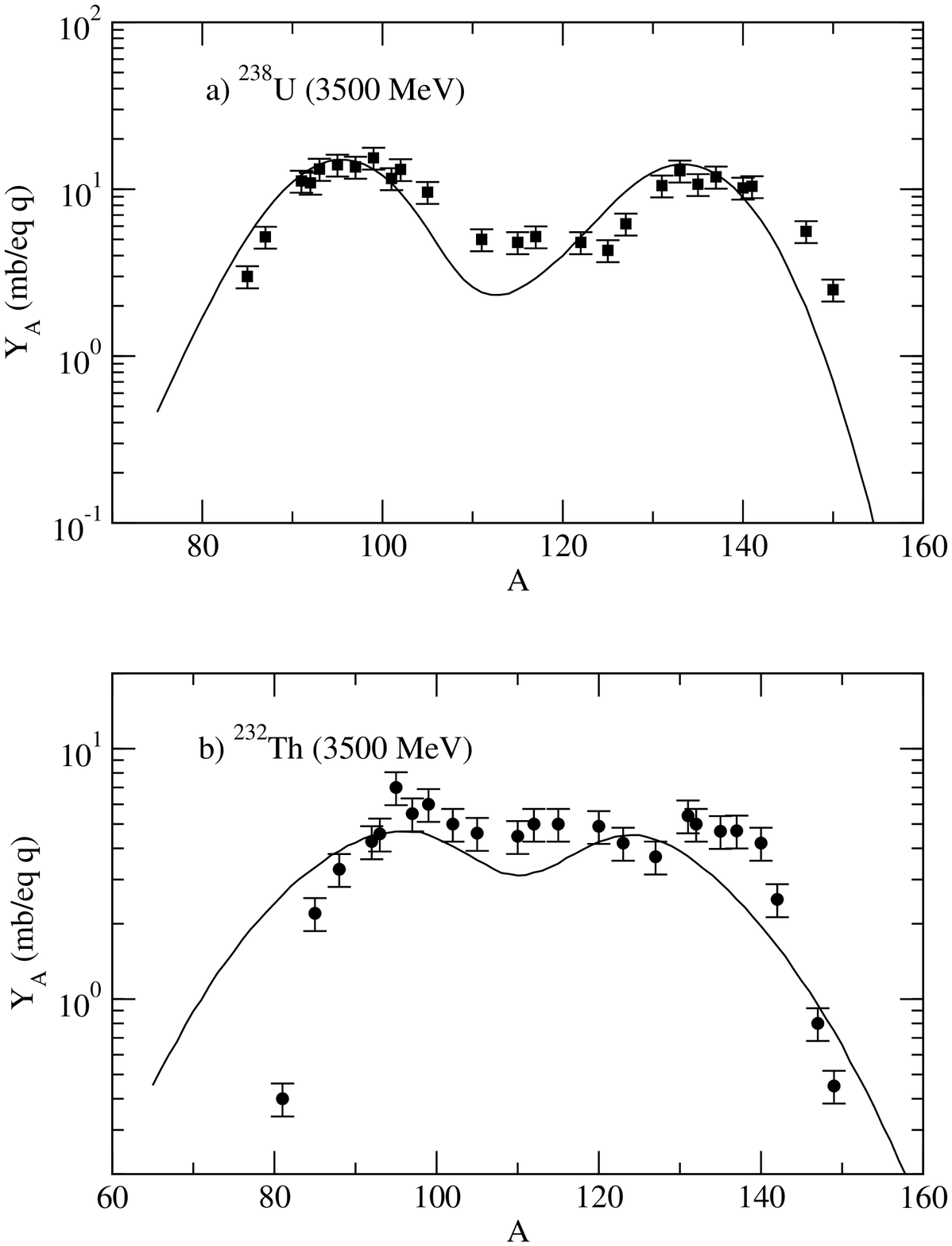,height=14cm,width=12cm,angle=0.}
  \caption{Mass distribution, mass-yield as a function of mass number
in units of mb per equivalent quanta (mb/eq q),
at $E_{max}=3500$ MeV for a) $^{238}$U and b) $^{232}$Th targets. 
The solid line is the result of CRISP calculation.}
\end{figure}

\begin{figure}
\epsfig{file=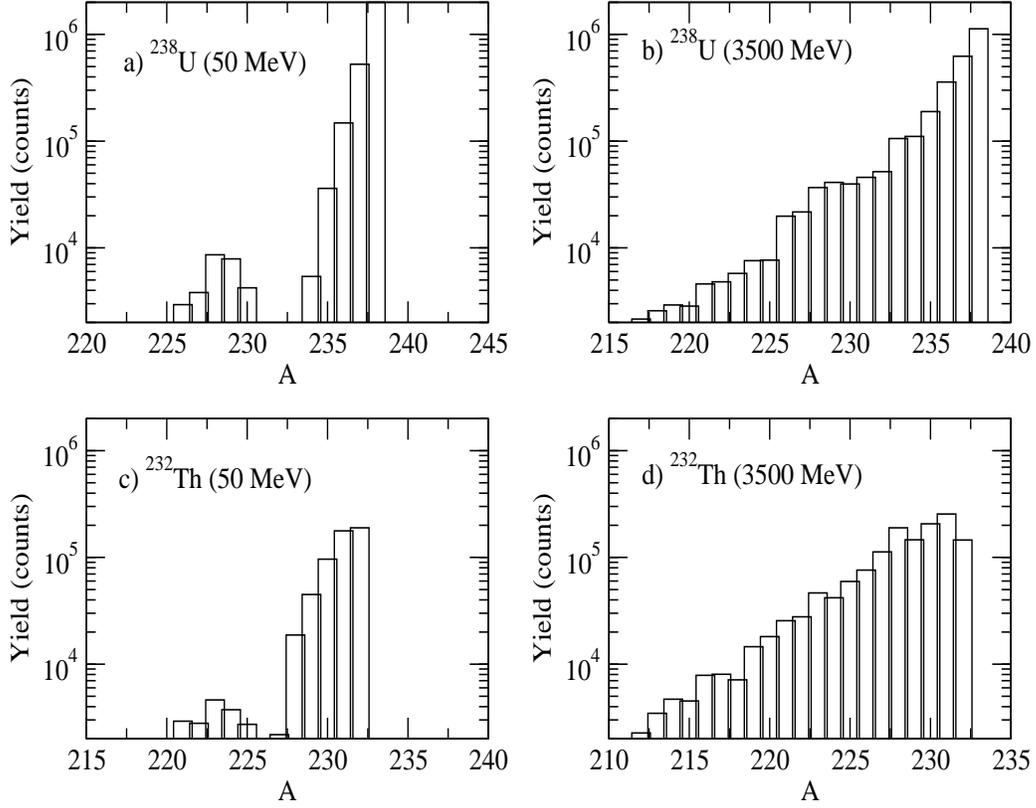,height=16cm,width=14cm,angle=-90.}
  \caption{Initial distribution for the fissioning nuclei for a,b) $^{238}$U 
and c,d) $^{232}$Th targets at bremsstrahlung endpoint energies 50 
and 3500 MeV, respectively.}
\end{figure}

\begin{figure}
\epsfig{file=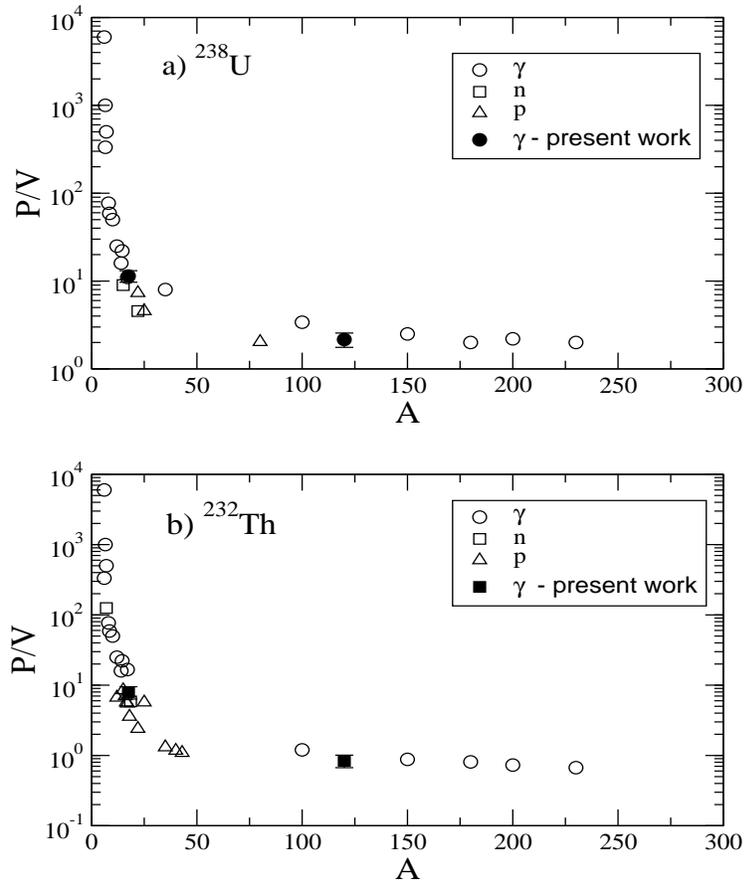,height=14cm,width=12cm,angle=0.}
\caption{The peak to valley ratio $P/V$ as a function of excitation energy
for fission induced by different probes as photon [42,43], 
protons [35,36,45] and neutrons [21,30,44] on a) ${238}$U and b) $^{232}$Th 
targets .}
\end{figure} 

\begin{figure}
\epsfig{file=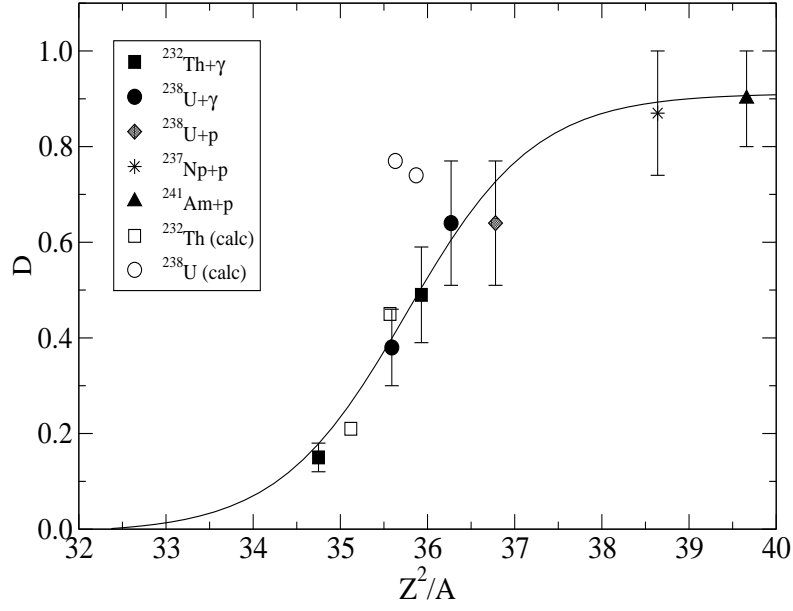,height=12cm,width=10cm,angle=-90.}
\caption{Fissility $D$ as a function of  $Z^{2}/A$ for
p+$^{237}$Np, p$^{238}$U, p+$^{241}$Am  \cite{Nina3, Nina4} , 
$\gamma$+$^{238}$U and $\gamma$+$^{232}$Th present work. 
Calculations by CRISP are open square and open circle symbols without 
error bars. The solid line is to guide the eye through the experimental points.}
\end{figure}

\end{document}